\documentclass[prd,showpacs,showkeys]{revtex4-2}
\usepackage{amsmath, amssymb}
\usepackage{hyperref}
\usepackage{graphicx}
\newcommand{\lr}[1]{\left(#1\right)}

\begin{document}
\title{Effects of a background scalar field induced by the Lorentz symmetry violation on Non Relativistic Quantum Mechanics}
\author{J. D. Garc\'ia-Aguilar}
\email{jdgarcia@ipn.mx}
\affiliation{Centro de Estudios Cient\'ificos y Tecnol\'ogicos No 16, Instituto Polit\'ecnico Nacional, Pachuca: Ciudad del Conocimiento y la Cultura, Carretera Pachuca Actopan km 1+500, San Agust\'in Tlaxiaca, Hidalgo, M\'exico.
}
\begin{abstract}
The extension of Standard Model of the Fundamental Particles which consider the Lorentz Symmetry Violation governed by a background tensor field  presented by Colladay and Kostelecky consider the hypothesis of having a privileged direction in the space-time. Here we investigated a extension of the Schrödinger Equation inspired on that model taking into account a background scalar field. We analyze the behavior of this modified Schrödinger equation on a particle trapped in a well and quantum harmonic oscillator.
\end{abstract}

\pacs{11.30.Cp, 11.30.Er, 12.60.-i, 03.65.Ge}

\keywords{Lorentz Violation, CPT Violation, Standard Model Extension, Bound states.}

\maketitle

\section{Introduction}
The Standard Model Extension which consider the CPT symmetry violation (SME) \cite{Colladay:1996iz, Colladay:1998fq} is a  proposal to explain some conflicts between experimental/observational data and the theoretical predictions of the Standard Model (SM) \cite{Moura:2020xdi, Aranda:2013cva,Aghababaei:2013dia,Brevik:2020cky,Antonelli:2020nhn,Crivellin:2020oov}. The SME assumes that all possible Lorentz violating terms arise from nonzero expectation values for Lorentz tensors which are considered as spacetime background fields in a fundamental theory. These background fields  give rise to preferred space time directions. 

Lorentz symmetry violation leads to some modification on particle properties such as energy-dependent dispersion relations and field equations (see, e.g., Refs. \cite{AmelinoCamelia:1997gz,Coleman:1997xq,Coleman:1998ti,Aloisio:2000cm}) its effects can make evident  considering the framework of SME in the ultrarelativistic and non relativistic limits. Currently, from the model building view, a large effort has been done in different branches of physics to find possibles signal on Lorentz symmetry violation, and therefore, provide experimental bounds on SME parameters. Some of these studios utilizing neutrinos \cite{Dighe:2008bu,Barenboim:2009ts,Datta:2003dg,Diaz:2016fqd, Hooper:2005jp}, photons \cite{Cambiaso:2012vb,Martinez-Huerta:2016odc, Colladay:2017zen,Malta:2019tgo}, kaons \cite{Vos:2014yqa}, harmonic oscillator under influence of a Coulomb-like potential \cite{Bakke:2012jy},  Bose-Einstein condensates \cite{Colladay:2006rt,Furtado:2020olp}, supersymmetry \cite{Berger:2001rm, Colladay:2010xf} and extra dimensions \cite{Rizzo:2005um,Garcia-Aguilar:2016kjy}. However, a complete treatment of the SME in the non-relativistic quantum mechanics sector remains an open issue.

In the present work, we seek to address this gap in the literature by extending the existing work of SME in non relativistic quantum mechanic sector to study its implications over Schrödinger equation, thereby opening the path for additional searching for Lorentz violation. To achieve this goal, we have investigated the effects of a background constant scalar field induced by  Lorentz symmetry breaking. Then, we search a bound  state solutions  considering a time independent modified Schrödinger equation and shown that the Lorentz violation terms can be eliminated through  a redefinition on wavefunctions.

This papper is organized as follows: in section \ref{freepart} we show how considering the CPT-odd sector of SME the momentum operator is modified and this leads a Schrödinger equation which is not invariant under parity transformation $x\rightarrow -x$, we also  determine the solution for this equation considering a free particle. In section \ref{boxcase}, we solve the modified Schrödinger equation for a particle trapped in a well whereas in section \ref{HOcase}  harmonic oscillator case is treated. Finally, in section \ref{conc} we present our conclusions.

\section{The modified Schrödinger equation}\label{freepart}
In order to obtain the modified Schrödinger equation, we perform the study considering the scalar CPT-odd sector of the Lorentz Violation Standard Model Extension \cite{Colladay:1996iz} whose Lagrangian is given by 
\begin{equation}
\mathcal{L}=\lr{D_\mu\phi}^\dagger D^\mu\phi+\mu^2\phi^\dagger\phi-\frac{\lambda}{3!}\lr{\phi^\dagger\phi}^2+ik^\mu\phi^\dagger D_\mu\phi+h.c., \label{LangOr}
\end{equation}
where $\phi$ denote the Higgs doublet, $D_\mu$ is the covariant derivative and $k^\mu$ is the coefficient which quantifies the Lorentz violation which has dimensions of mass and obeys the relation $k_\mu k^\mu\ll 1$. 

For our purposes we are going to consider a massless scalar field without interactions. Therefore, the equation \ref{LangOr} is reduced to 
\begin{equation}
\mathcal{L}=\partial_\mu \phi^\dagger \partial^\mu \phi+i k_\mu \phi^\dagger\partial^\mu\phi-i k^\mu\phi\partial_\mu\phi^\dagger,
\end{equation} 
which can be rewritten as
\begin{equation}
\mathcal{L}=\partial'_\mu\phi^\dagger\partial'^\mu\phi,
\end{equation}
considering
\begin{equation}
\partial'_\mu=\partial_\mu+i k_ \mu, \label{newderiv}
\end{equation}
so, the redefinition for the derivative operator allows to eliminate the CPT violating term. 

Hence, the redefinition \ref{newderiv} establish the possibility to investigate the effects of the momentum operator redefinition 
given by
\begin{equation}
\hat{p}=-i\hbar\lr{\frac{d}{dx}+i\alpha} \label{modmom}
\end{equation}
where $\alpha$ is a background constant scalar field whose value quantifies the effects of  Lorentz symmetry violation.

The equation \ref{modmom} leads to rewrite the Schrödinger equation for the free particle in the form $\lr{\hbar=1}$
\begin{equation}
-\frac{1}{2m}\lr{\frac{d^2\psi}{dx^2}+2i\alpha\frac{d\psi}{dx}-\alpha^2\psi}=E\psi, \label{SchoUno}
\end{equation}
where,  it is straightforward to see how the equation \ref{SchoUno} is not invariant under $x\rightarrow -x$ parity transformation.

We can rewrite the equation \ref{SchoUno} in the form 
\begin{equation}
\frac{d^2\psi}{dx^2}+2i\alpha\frac{d\psi}{dx}=-\lr{k^2-\alpha^2}\psi, 
\end{equation}
where we define $k=\sqrt{2mE}$.

By imposing that $k^2>\alpha^2$, we obtain the solution
\begin{equation}
\psi\lr{x}=Ae^{i\lr{k-\alpha}x}+Be^{-i\lr{k+\alpha}x}, \label{sol1}
\end{equation}
where, we can see how the effects of Lorentz symmetry violation is present on wavefunctions whereas energy eigenvalues is independent of background scalar field.

Observe that in the limit $ \alpha \rightarrow 0 $ the Lorentz symmetry is restored,  so the equations \ref{modmom}, \ref{SchoUno}  and \ref{sol1} are analogous to the usual equations to describe the quantum free particle \cite{Landau:1991wop,Griffiths:1987tj}. This is to be expected because if we make a wavefunction redefinition 
\begin{equation}
\psi\lr{x}\rightarrow e^{-i\alpha x}\psi\lr{x},
\end{equation}
then the equation \ref{SchoUno} becomes
\begin{equation}
-\frac{1}{2m}\frac{d^2\psi}{dx^2}=E\psi,
\end{equation}
which is the canonical Schrödinger equation for the free particle.

\section{Particle on a box}\label{boxcase}
In this section, we investigate the effects of the background constant constant field over a particle trapped in a well with potential given by
\begin{equation}
V\lr{x}=\left\lbrace \begin{array}{lr}
0& 0\leq x\leq L\\
\infty& \mbox{otherwise}
\end{array}\right. \label{pot}
\end{equation}
and hence a particle is completely free, except at the two ends and therefore the wavefunction is only different zero inside the well.

We know from section \ref{freepart} that the solution for the free particle considering the operator \ref{modmom} (in order to consider the Lorentz symmetry violation) and subject to the condition
\begin{equation}
\psi\lr{0}=\psi\lr{L}=0,
\end{equation}
is given by
\begin{equation}
\psi_n\lr{x}=Ae^{-i\alpha x}\sin\lr{k_nx}, \label{pozosol}
\end{equation}
where
\begin{equation}
k_n=\frac{n\pi}{L}
\end{equation}
with $n=1,2,3,\ldots$.

From the expression \ref{pozosol} we obtain the eigenvalues for the Energy in the form
\begin{equation}
E_n=\frac{n^2\pi^2}{2mL^2}
\end{equation}
and the normalization constant sets
\begin{equation}
A=\sqrt{\frac{2}{L}}.
\end{equation}

\begin{figure}[h]
\includegraphics[scale=0.7]{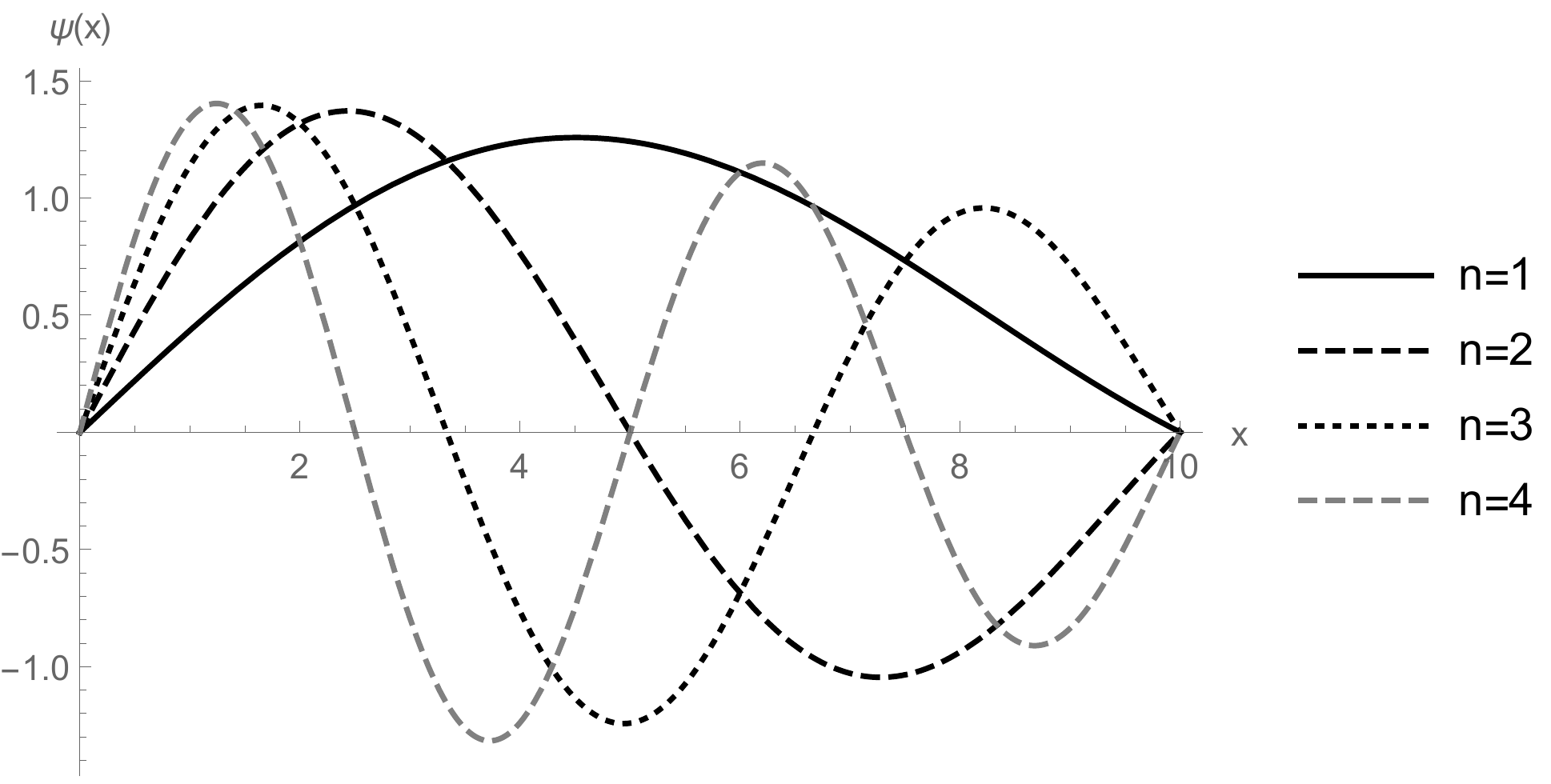}
\caption{First four plots of the real parts of $\psi_n\lr{x}$ for $\alpha=1/10$ and $L=10$.}
\label{graficapozo}
\end{figure}

Assuming that $\alpha=1/10$ and $L=10$ for purposes of demonstration we obtain the plot of the first four graphs of the real part to the equation \ref{pozosol} shown in figure \ref{graficapozo}; here we see the dependence of wavefunctions on $\alpha$ value. As we expected in the limit $\alpha\rightarrow 0 $, the solution have the same features when  is taking into account the usual definition for the momentum operator on the system \cite{Landau:1991wop,Griffiths:1987tj}.

We have know an important result: the  effects of Lorentz symmetry violation appears only over wave functions for this system, the observables are independent of background scalar field. Therefore, the Lorentz Violation does not lead to alterations on the Heisenberg uncertainty principle \cite{Schrodinger:1930ty}.

\section{The Harmonic Oscillator case}\label{HOcase}
In this section, our aim is to solve the Schrödinger equation taking into account  the operator \ref{modmom} and the potential given by
\begin{equation}
V=\frac{1}{2}m\omega^2x^2.
\end{equation}

Thus, the time independent Schrödinger equation to solve is given in the form
\begin{equation}
-\frac{1}{2m}\lr{\frac{d^2\psi}{dx^2}+2i\alpha\frac{d\psi}{dx}-\alpha^2\psi}+\frac{1}{2}m\omega^2 x^2\psi=E\psi \label{oscieq}.
\end{equation}
which we can solve considering the following ladder operators
\begin{equation}
a_\pm=\frac{1}{\sqrt{2m\omega}}\lr{\mp\lr{\frac{d}{dx}+i\alpha}+m\omega x}, \label{ladder}
\end{equation}
where direct calculation reveals that 
\begin{equation}
\left[a_{-},a_{+}\right]=1.
\end{equation}

With this, equation \ref{oscieq} becomes
\begin{equation}
\omega\lr{a_{+}a_{-}+\frac{1}{2}}\psi=E\psi,
\end{equation}
and through the annihilation operator we can determine the ground state ($\psi_0$) of the system
\begin{equation}
a_{-}\psi_0=\frac{1}{\sqrt{2m\omega}}\lr{\lr{\frac{d}{dx}+i\alpha}+m\omega x}\psi_0=0.
\end{equation}

The solution for this differential equation is straightforward to obtain and is given by
\begin{equation}
\psi_0\lr{x}=Ae^{-\frac{m\omega}{2}x^2}e^{-i\alpha x} \label{solground}
\end{equation}
where the normalization constant take the value
\begin{equation}
A=\lr{\frac{m\omega}{\pi}}^{1/4},
\end{equation}
and to determine the energy  we substitute the equation \ref{solground}  into right side of the equation \ref{oscieq}, then
\begin{equation}
E_0=\frac{\omega}{2},
\end{equation}
we can see how the last two results does not depend on the $\alpha$ parameter.

To generate the excited states we can use the creation operator $a_{+}$ over ground state repeatedly
\begin{equation}
\psi_n\lr{x}=A_n a_{+} \psi_0\lr{x}, \label{staex}
\end{equation}
where $A_n$ is the normalization constant and the energy is incremented by $\omega$ with each step
\begin{equation}
E_n=\lr{n+\frac{1}{2}}\omega.
\end{equation}

\begin{figure}[h]
\includegraphics[scale=0.7]{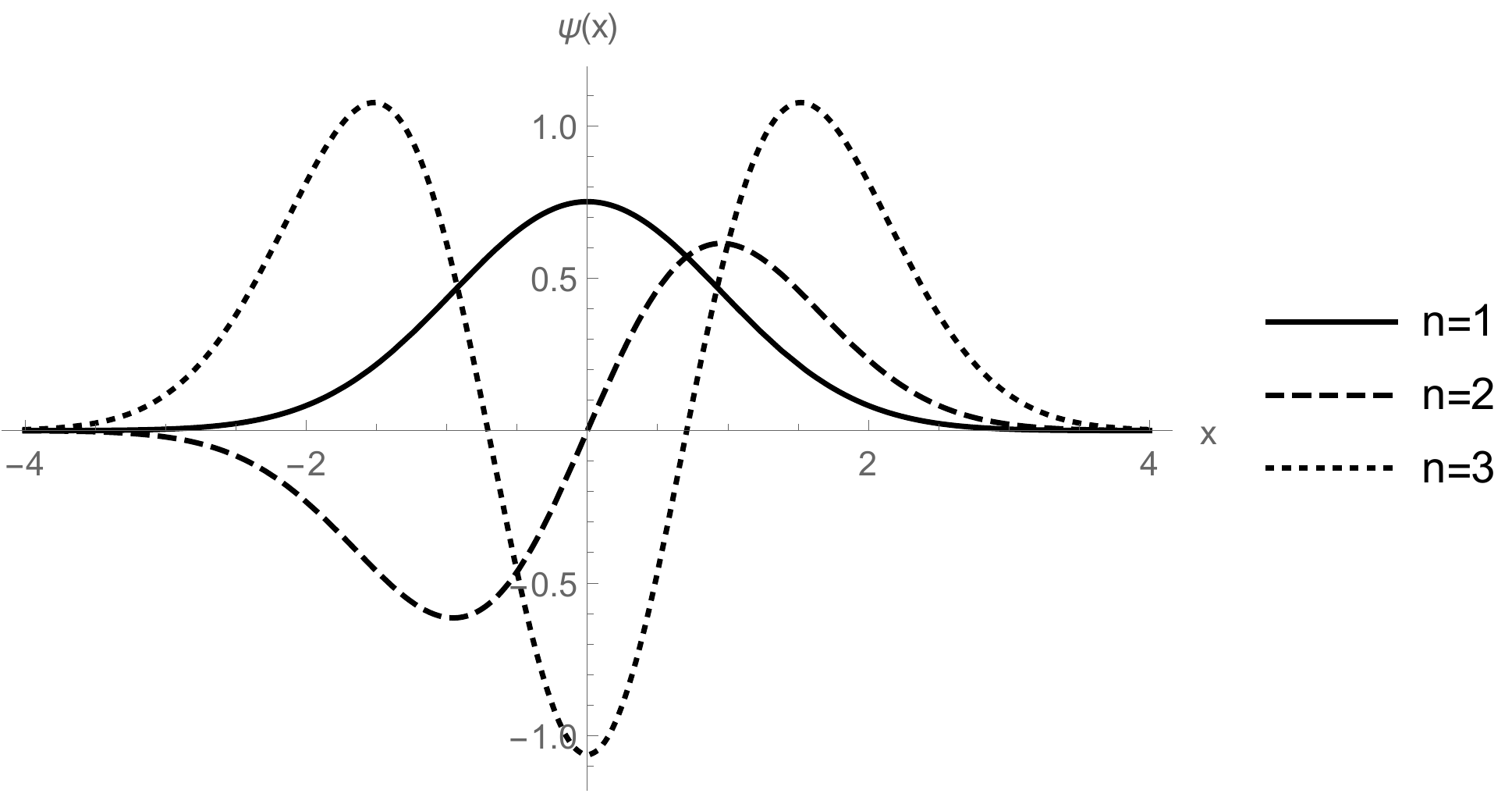}
\caption{First three plots of the real parts of $\psi_n\lr{x}$ considering $\alpha=1/10$ and $m=\omega=1$.}
\label{graficaoscilador}
\end{figure}

\begin{figure}[h]
\includegraphics[scale=0.7]{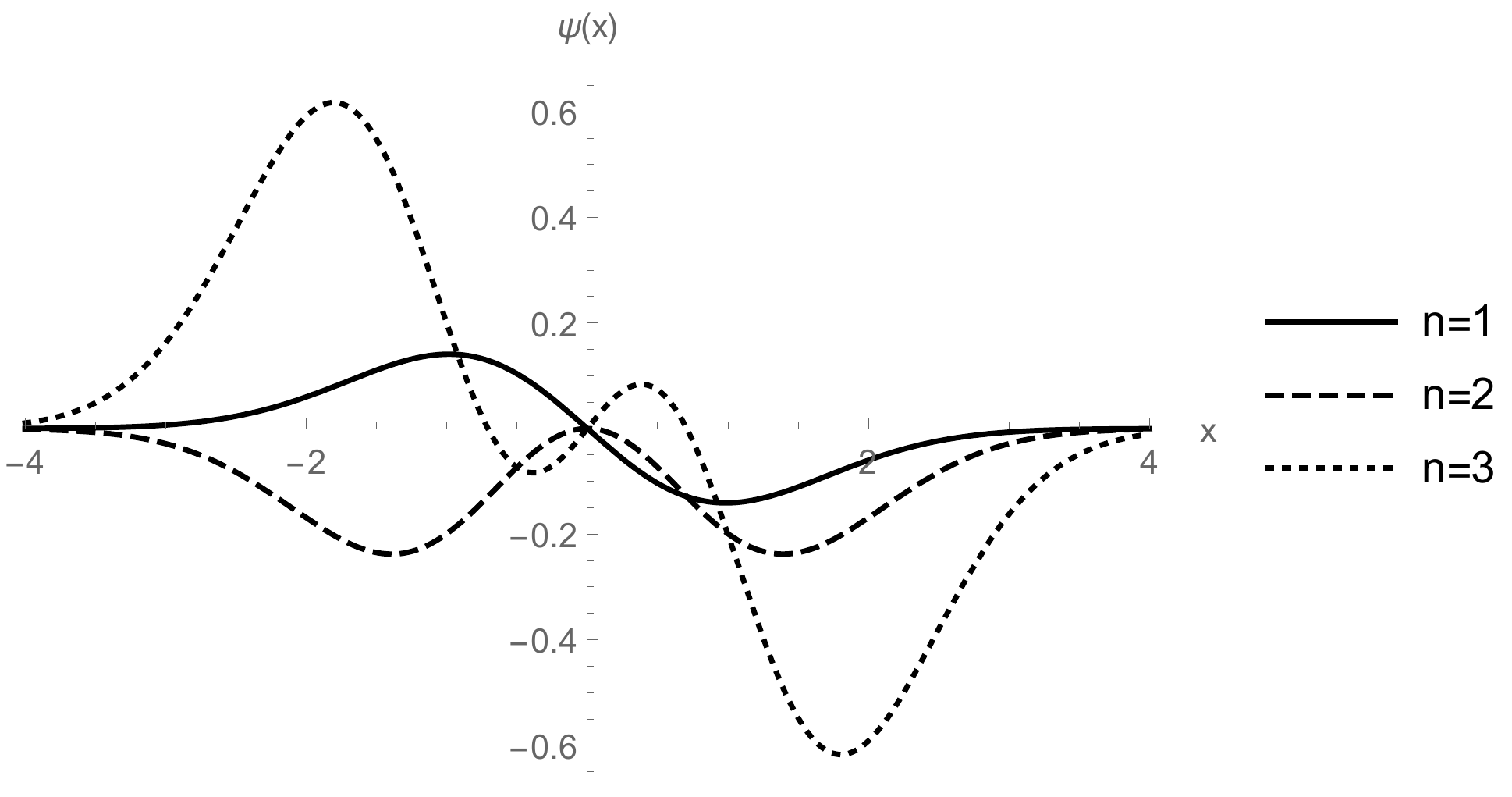}
\caption{First three plots of the imaginary parts of $\psi_n\lr{x}$ considering $\alpha=1/10$ and $m=\omega=1$.}
\label{graficaosciladorim}
\end{figure}

In figure \ref{graficaoscilador} we plot the first three graphs of real part of the wavefunctions given by the equation \ref{staex} using the equivalence showed in appendix \ref{primap} considering $m=\omega=1$ and $\alpha=10$. In figure \ref{graficaosciladorim} we can see explicitly the effects of Lorentz violation over wave functions because the imaginary part of them it is different to zero .

In this section we have seen that  similar to the previous section, the effects of Lorentz symmetry violation is presented only over the wave functions, the energy levels are independent on $\alpha$ parameter. 

\section{Conclusions}\label{conc}
This papper deals with time independent Schrödinger equation under effect of the CPT-odd Lorentz violation. We have shown that momemtum operator is modified and, consequently provides a modified Schrödinger equation to be investigated. In this sense, we present the wavefunctions for two different systems.

In our first analysis, we considered a particle trapped in a well, where we have shown  that the effects of the Lorentz violation appears only on wavefunctions whereas the energy levels and normalization constant are similar to the case where Lorentz symmetry is invariant. In the second analysis the harmonic oscilator is solved  through a redefinition on the ladder operators and we find that similarly  to the first analysis, the Lorentz violation is present only on the wavefunctions.

As we seen the presence of Lorentz symmetry violation  in the Schrödinger equation for a particle in a well and harmonic oscillator  can not induce modifications over energy levels of the system and therefore, it is not possible to expect signal of SME in this energy scale. It is worth mentioning that our analysis can be considered beyond one dimensión and it is in our interest as future perspective to analize the two and three dimensional case.

\begin{acknowledgments}
We would like to thank Pablo Paniagua López for their comments and discussión on the manuscript. The author appreciate the facilities given by IPN through the SIP project number 20201313. This work was partially supported by a PAPIIT grant IN109321.
\end{acknowledgments} 

\appendix

\section{Equivalence formula}\label{primap}
It is to interesting to consider what happens when some redefinitions over the wavefunctions is used before applies the ladder operators of the equation \ref{ladder} . Since the ground state give by the equation \ref{solground} can be written as
\begin{equation}
\psi_0\lr{x}=e^{-i\alpha x} \psi'_0\lr{x}
\end{equation}
where  
\begin{equation}
\psi'_0\lr{x}=\lr{\frac{m\omega}{\pi}}^{1/4}e^{-\frac{m\omega}{2}x^2},
\end{equation}
is the usual ground state for quantum oscillator where Lorentz symmetry is present.

Through algebraic manipulations
\begin{eqnarray}
\psi_1\lr{x}&=&A_1a_{+}\psi_0\lr{x}\nonumber\\
&=&\frac{A_1}{\sqrt{2m\omega}}\lr{-\lr{\frac{d}{dx}+i\alpha}+m\omega x}e^{-i\alpha x} \psi'_0\lr{x}\nonumber\\
&=&e^{-i\alpha x}\lr{\frac{A_1}{\sqrt{2m\omega}}\lr{-\frac{d}{dx}+m\omega x} \psi'_0\lr{x}}\\
&=&e^{-i\alpha x}\psi'_1\lr{x},\nonumber
\end{eqnarray}
where $\psi'_1\lr{x}$ is the first excited state for usual quantum oscillator \cite{Landau:1991wop,Griffiths:1987tj}. 

Then, we can determine by similar way the general equation for the excited states 
\begin{equation}
\psi_n\lr{x}=e^{-i\alpha x} \psi'_n\lr{x}.
\end{equation}

\end{document}